\documentclass[12pt,leqno]{article}
\catcode`\@=11 \long\def\@makecaption#1#2{
 \vskip 10pt
 \setbox\@tempboxa\hbox{\bf #1: \sf #2}
 \ifdim \wd\@tempboxa >\hsize \bf #1: \sf #2\par \else \hbox
to\hsize{\hfil\box\@tempboxa\hfil}
 \fi}
\catcode`\@=12
\newfont{\Bbb}{msbm10 scaled\magstep1}
\newcommand{\R}{\mbox{\Bbb R}}

\newcommand{\conv}{{\rm conv}}

\newtheorem{theo}{Theorem}
\newtheorem{lem}[theo]{Lemma}

\newenvironment{proof}[1][Proof]{\textbf{#1.} }{\mbox{}\hfill \rule{0.5em}{0.5em}}

\begin{document}
\title{A note on curves equipartition}
\author{
Mario A. Lopez\\ {\small Department of Computer Science}\\ {\small
University of Denver}\\ {\small Denver, CO 80208, U. S. A.}\\
{\small \tt mlopez@cs.du.edu} \and
Shlomo Reisner\\ {\small Department of Mathematics}\\
{\small University of Haifa}\\ {\small Haifa 31905, Israel}\\
{\small \tt reisner@math.haifa.ac.il}}
\date{}
\maketitle

\noindent
The problem of the existence of an equi-partition of a curve in $\R^n$
has recently been raised in the context of computational geometry (see
\cite{PT} and \cite{PGT}). The problem is to show that for a
(continuous) curve $\Gamma\,:\,[0,1]\rightarrow \R^n$ and for any positive
integer $N$, there exist points $t_0=0<t_1<\ldots<t_{N-1}<1=t_N$, such that
$d(\Gamma(t_{i-1}),\Gamma(t_i))=d(\Gamma(t_{i}),\Gamma(t_{i+1}))$ for
all $i=1,\ldots ,N$, where $d$ is a metric or even a semi-metric (a weaker
notion) on $\R^n$.

In fact, this problem, for $\R^n$ replaced by any metric space $(X,d)$ was
given a positive solution by Urbanik \cite{Urbanik} under the (necessary)
constraint $\Gamma(0)\neq\Gamma(1)$.

We show here that the existence of such points, in a broader
context, is a consequence of Brower's fixed point theorem (for
reference see any intermediate level book on topology, e.g.\
\cite{Dugundji}). \vspace{5mm}

Let $\Delta^n$ be the $n$-dimensional simplex and
$\sigma^n$ be its boundary. Let $\tau$ be a permutation of
$\{0,\ldots,n\}$. We define a map
$\varphi_{\tau}\,:\,\Delta^n\rightarrow \Delta^n$ in the following
way:

\noindent
Every $x\in\Delta^n$ has a unique representation $x=\sum_{i=0}^n\alpha_i(x)
p_i$, where $p_i$ are the vertices of $\Delta^n$ and $0\leq \alpha_i(x)\leq 1$,
$\sum_{i=0}^n\alpha_i(x)=1$. We then define:
\[\varphi_{\tau}(x)=\sum_{i=0}^n\alpha_i(x)p_{\tau(i)}\,.\]
$\varphi_{\tau}$ is, of course, continuous.

\begin{lem}
\label{lem-A}
Let $\tau$ be a cyclic permutation of $\{0,1,\ldots,n\}$ and let
$\Psi\,:\,\sigma^n\rightarrow \sigma^n$ be a map such that, for every
proper face $A$ of $\Delta^n$, $\Psi(A)\subset \varphi_{\tau}(A)$. Then
$\Psi$ has no fixed point.
\end{lem}
\begin{proof}
Assume that $\Psi(x)=x$ for some $x\in \sigma^n$ and let $x=\sum_{i=0}^n\alpha_i(x)
p_i$ be the representation of $x$. Let $A={\rm face}(x)$ be the minimal face of
$\Delta^n$ containing $x$. Without loss of generality we may assume that
$A=\conv\{p_0,\ldots,p_k\}$ for some $0\leq k<n$. That is, $x=\sum_{i=0}^k\alpha_i(x)
p_i$ and $\alpha_i(x)>0$ for $i=0,\ldots,k$. Now, by the assumption,
$\Psi(x)\in \varphi_{\tau}(A)$, thus $x=\Psi(x)= \sum_{i=0}^k \beta_i p_{\tau(i)}$.
By the uniqueness of the representation we have: $\{p_{\tau(0)},\ldots,p_{\tau(k)}\}
=\{p_0,\ldots,p_k\}$ (and $\beta_i=\alpha_{\tau^{-1}(i)}$ for all $0\leq i\leq k$).
But, since $\tau$ is cyclic, no proper subset of $\{0,\ldots,n\}$ is mapped by
$\tau$ on itself. This is a contradiction.
\end{proof}
\vspace{2mm}\mbox{}

Let $X$ be a non-empty set and $\Gamma\,:\,[0,1]\rightarrow X$ a function. Let
$d\,:\,X\times X\rightarrow \R_+$ (that is $d(x,y)\geq 0$ for all $x,y\in X$)
satisfy $d(x,x)=0$.
Assume also that $d(\Gamma(s),\Gamma(t))$
is continuous on $[0,1]\times [0,1]$ (this last property holds, for example,
if $X$ is a metric space, $\Gamma$ is continuous and $d$ is continuous on
$X\times X$).
Associated with $d$ as above, we denote, for
$t_1,t_2\in [0,1]$, $D(t_1,t_2)=d(\Gamma(t_1),\Gamma(t_2))$.

Theorem~\ref{th-B}
which follows is motivated by the above context, but, as stated, is true for
any continuous function $D\,:\,[0,1]\times [0,1]\rightarrow \R_+$. In the particular case
that $D$ is constructed as above, with $d$ a metric on $X$, Theorem~\ref{th-B} implies
Urbanik's result \cite{Urbanik}.

\begin{theo}
\label{th-B}
Given a continuous function $D\,:\,[0,1]\times [0,1]\rightarrow \R_+$
such that $D(t,t)=0$ for all $t\in [0,1]$. Then
for every positive integer $N$
there exist points $t_0=0\leq t_1\leq\cdots \leq t_{N-1}\leq1=t_N$ such that
$D(t_{i-1},t_i)=D(t_{i},t_{i+1})$ for $i=1,\ldots,N-1$.

\noindent
Moreover, if for a given $N$ there are no points $t_0=0\leq t_1\leq \cdots \leq t_{N-1}\leq 1=t_N$
such that $D(t_{i-1},t_i)=0$ for $i=1,\ldots,N-1$, then
for every sequence
of positive real numbers $\alpha_1,\ldots,\alpha_N$ such that $\sum_{i=1}^N \alpha_i=1$,
there exist points $t_0=0<t_1<\cdots <t_{N-1}<1=t_N$ such that
\[\frac{D(t_{i-1},t_i)}{D(t_{i},t_{i+1})}=\frac{\alpha_i}{\alpha_{i+1}}
\mbox{\quad for $i=1,\ldots,N-1$}\,.\]
\end{theo}
\begin{proof}
If there are points $t_0=0\leq t_1\leq \cdots \leq t_{N-1}\leq 1=t_N$
such that $D(t_{i-1},t_i)=0$ for $i=1,\ldots,N-1$, then the first claim is obvious.
Thus let us assume that such a sequence does not exist and prove the ``Moreover''
assertion.

Let $S$ be the $(N-1)$-dimensional simplex
\[S=\{(t_1,\ldots,t_{N-1})\,|\,0\leq t_1\leq\cdots\leq t_{N-1}\leq 1\}\]
and let $F\,:\,S\rightarrow \R^N$ be defined by
\[F(t_1,\ldots,t_{N-1})=(D(0,t_1),D(t_1,t_2),\ldots,D(t_{N-1},1))\,.\]
Then, $F$ is continuous and $F(S)\subset \R^N_+$. Moreover, every proper face
$A$ of $S$ is characterized as being the set of $(N-1)$-tuples $(t_1,\ldots,t_{N-1})\in S$
such that, in the chain of inequalities $0\leq t_1 \leq \cdots \leq t_{N-1} \leq 1$, there
are $0<k\leq N-1$ equalities (the dimension of $A$ is then $N-1-k$). Thus, for
$0<k\leq N-1$ and a $(N-1-k)$-dimensional face $A$ of $S$, $F(A)$ is contained in the
subspace $\cal A$ of $\R^n$:
\[{\cal A}=\{(x_1,\ldots,x_N)\,|\, x_i=0 \mbox{\ whenever\ } t_{i-1}=t_i \mbox{\ for all\ }
(t_1,\ldots,t_{N-1})\in A\}\]
(here, and in the sequel, $t_0=0$ and $t_N=1$).\vspace{1mm}

Let $\Sigma$ be the $(N-1)$-dimensional simplex which is the convex hull in $\R^N$ of
the unit coordinate vectors $e_i$, $i=1,\ldots,N$.
Let $G\,:\,\R^N_+\setminus \{\overline{0}\}\rightarrow \Sigma$
be the radial projection with center $\overline{0}\in \R^N$. The composed map $GF$ maps $S$
continuously into
$\Sigma$ (as we have excluded the possibility that $\overline{0}\in F(S)$).\vspace{1mm}

\noindent
{\bf Claim}\quad {\em The relative interior of $\Sigma$ is contained in $GF(S)$.}\vspace{1mm}

\noindent
We remark that, by simple induction, it follows from the above Claim that, in fact, $GF$ maps
$S$ {\em onto\/} $\Sigma$, but the Claim as is, is clearly sufficient to complete the proof of
Theorem~\ref{th-B}, since
it implies that, for any positive $\alpha_1,\ldots,\alpha_N$ with $\sum_{i=1}^N\alpha_i=1$,
the ray $L$ from the origin in $\R^N$ with parametric equation
\[L\,:\,\overline{x}=(\alpha_1t,\ldots,\alpha_Nt),\;\;t>0\,,\]
that intersects $\Sigma$ at the point $(\alpha_1,\ldots,\alpha_N)$, must, by the Claim,
intersect also $F(S)$.\vspace{1mm}

\noindent
{\bf Proof of the Claim.}\quad Assume, for contradiction, that there exists a point $P$ in
the relative interior of $\Sigma$, such that $P\not\in GF(S)$. Let
\[H\,:\,\Sigma\setminus \{P\}\rightarrow \partial\Sigma\]
be the radial projection onto $\partial\Sigma$ with center $P$ (in the hyperplane containing $\Sigma$).
$H$ is continuous on $\Sigma\setminus \{P\}$ and, since $P\not\in GF(S)$, the map
$HGF\,:\,S\rightarrow \partial\Sigma$ is continuous. $H$ restricted to $\partial\Sigma$ is the
identity, thus, by the previous discussion, $HGF(A)\subset\Sigma\cap{\cal A}$ for every proper face
$A$ of $S$. Let us identify now $S$ with $\Sigma$ by the natural affine homeomorphism that
identifies every face $A$ of $S$ with $\Sigma\cap {\cal A}$. Via this identification, $HGF$
induces a continuous map
\[\Phi\,:\,\Sigma\rightarrow \partial\Sigma\] with the property that $\Phi(A)\subset A$ for
every proper face $A$ of $\Sigma$.

Now let $\varphi_{\tau}\,:\,\Sigma\rightarrow\Sigma$ be the map associated with a cyclic permutation
of the vertices of $\Sigma$, as in the context of Lemma~\ref{lem-A}, and let
\[\Psi=\varphi_{\tau}\Phi\,:\,\Sigma\rightarrow \partial\Sigma\,.\]

\noindent
Then $\Psi$ restricted to $\partial\Sigma$ satisfies the assumptions of Lemma~\ref{lem-A}, hence
$\Psi$ has no fixed point in $\partial\Sigma$. Considering $\Psi$ as a map from $\Sigma$ into
$\Sigma$ we get a continuous map from $\Sigma$ into itself that has no fixed point. This contradicts
Brower's fixed point theorem.
\end{proof}
\vspace{2mm}\mbox{}

\noindent
{\bf Reminder:}\quad Brower's fixed point theorem states that any continuous map from
$B^{K}$ into $B^{K}$ has a fixed point ($B^{K}$ is the closed unit ball of $\R^K$ but,
of course, can be replaced by anything homeomorphic to it, like $\Sigma$ here).

\noindent
{\bf Acknowledgement:}\quad The authors thank Jim Hagler for
very important comments.

\end{document}